\documentclass[10pt]{article}
\date{}
\title{Qualms regarding \lq\lq Superstatistics \rq\rq\
by C. Beck and E. G. D. Cohen, \emph{Physica A} \textbf{321} (2003) (cond-mat/0205097)}
   \author{B. H. Lavenda$^1$ and J. Dunning-Davies$^2$\\
$^1$Universit\`a degli Studi  Camerino 62032 (MC) Italy;\\ email: bernard.lavenda@unicam.it\\
$^2$ Department of Physics, University of Hull, Hull HU6
7RX\\ England; email: j.dunning-davies@hull.ac.uk}
\usepackage{eufrak}

\newcommand{\threehalves}{\mbox{\small$\frac{3}{2}$}}

\newcommand{\ebar}{\bar{E}}

\begin{document}
\maketitle
\begin{abstract} 
There is neither motivation nor need to introduce \lq superstatistics.\rq
\end{abstract}
\flushbottom
The authors attempt to generalize the Boltzmann factor so as to obtain a more general
statistics, \emph{i.e.\/}, their so-called superstatistics. The do so by performing
a Laplace transform on the probability density function (pdf) of an intensive variable, $f(\beta)$, where $\beta$
is the inverse temperature. The criteria they use for choosing $f(\beta)$ are the
following:
\begin{enumerate}
\item It must be normalized
\begin{equation}
\int_0^\infty\,f(\beta)\,d\beta=1. \label{eq:norm}
\end{equation}
\item It must be such that its Laplace transform
\[
B(E)=\int_0^\infty\, e^{-\beta E}f(\beta)\,d\beta \]
is normalized, or at least is normalized with respect to a density of states $\rho(E)$
\[
\int_0^\infty\,B(E)\rho(E)\,dE=\int_0^\infty\,d\beta f(\beta)\int_0^\infty\,dE
e^{-\beta E}\rho(E)=1. \]
\item
The superstatistics should reduce to BG-statistics when there are no fluctuations in
the intensive variables.
\end{enumerate}
In thermodynamics there are two types of variables: \emph{extensive\/} and 
\emph{intensive} variables. In mathematical statistics they correspond to 
\emph{observable\/} and \emph{estimable\/} variables, respectively 
\cite{Mandelbrot,Lavenda}. Estimable variables  relate the probability distribution
of the extensive variables to the properties of the physical system. They are referred
to as the \lq state of nature\rq\ in mathematical statistics. Estimable variables change
due to the nature of the physical interaction, and, therefore, cannot be considered random
variables in the frequency sense. Estimable variables cannot be treated in the 
limit-of-frequency sense; rather, it must be interpreted in the subjective sense that some values
of $\beta$ are more \lq probable\rq\ than others.\par
If the parameter's value is completely unknown in so far as it can take on any conceivable
value from $0$ to $\infty$, Jeffreys' rule \cite{Jeffreys} for choosing the prior is to take its logarithm
as uniform so that
\begin{equation}
f(\beta)\,d\beta\propto\frac{d\beta}{\beta}\;\;\;\;\;\;\;\;\; 0<\beta<\infty.
\label{eq:Jeffreys}
\end{equation}
Since the integral of the prior is infinite, it is called improper. Jeffreys uses $\infty$
to represent the probability of a certain event rather than $1$. The indetermancy for
predicting whether will fall within the interval $(0,b)$, $\int_0^b\,d\beta/\beta=
\infty$, or $(b,\infty)$, $\int_b^\infty\,d\beta/\beta=\infty$, is merely a formal 
representation of ignorance.\par
However, Beck and Cohen do not imply that their $f(\beta)$ is a prior pdf. Transforming an improper pdf into a proper pdf can be accomplished
by use of Bayes' theorem \cite{Zellner}. The method uses a likelihood function which 
essentially inverts the roles of the observable and estimable variables. Due to the
additivity of the observable variable, the likelihood function will be a product of the
individual probabilities. The likelihood function takes us from an improper, 
uninformative, pdf to a  proper, informative, pdf. This is undoubtedly the meaning of 
$f(\beta)$.\par
Bayes' theorem can be written as \cite{Lavenda}
\begin{equation}
f(\beta;\ebar)=\frac{e^{\mathcal{L}(\beta:\ebar)}}{B(\ebar)}f(\beta) 
\label{eq:Bayes}
\end{equation}
where $\mathcal{L}(\beta;\ebar)$ is the log-likehood function which transforms the
prior pdf $f(\beta)$ into the posterior pdf $f(\beta;\ebar)$, where $\ebar$ is the
average energy of the sample. Since the posterior pdf is normalized, we obtain
the explicit expression for $B(\bar{E})$ as
\[
B(\ebar)=\int_0^\infty\, e^{-\beta\ebar-L(\beta)}f(\beta)\,d\beta=
\int_0^\infty\, e^{-\beta\ebar-L(\beta)}\frac{d\beta}{\beta}, \]
where $L(\beta)$ is the Legendre transform of the entropy with respect to the energy, and
we have introduced Jeffreys' improper prior pdf into the second expression.
\par
We now turn to the examples treated by Beck and Cohen. According to the authors, the
most relevant example of superstatistics is the Gamma pdf
\begin{equation}
f(\beta)=\frac{1}{b\Gamma(c)}\left(\frac{\beta}{b}\right)^{c-1}e^{-\beta/b}. 
\label{eq:G}
\end{equation}
This Gamma pdf for the inverse temperature has already been derived in ref 
\cite{Lavenda} eqn (4.97) from Bayes' theorem (\ref{eq:Bayes}),  where $2c=3N$ 
is the effective number of degrees-of-freedom and 
$b=1/E_0$ is a fixed parameter. Setting the authors' $f(\beta)=e^{-L(\beta)}/\beta$, 
we find
\[
L(\beta)=\beta/b-c\log(\beta/b)+\log\Gamma(c) \]
Since $L(\beta)$ is the Legendre transform of the entropy, it is the logarithm of the
partition function so that
\[
\frac{\partial L}{\partial\beta}=\frac{1}{b}-\frac{c}{\beta}=-\ebar \]
which gives the thermal equation of state
\begin{equation}
\beta=\frac{\beta_0}{1+b\ebar},\label{eq:thermal}
\end{equation}
where $\beta_0=bc=\threehalves N/E_0$, in our notation. The entropy is the Legendre 
transform of $L$ so that for $c$ large enough so that Stirling's approximation holds we 
get
\begin{equation}
S(E)=c\log(1+b\ebar). \label{eq:S}
\end{equation}
This means that
\[
B(\ebar)=e^{-S(\ebar)}. \]
\par
It is imperative to emphasize that absolutely no reference to superstatistics 
has been made! Fluctuations in intensive variables occur in BG-statistics. In fact for
$b\ebar\gg1$ the thermal equation of state (\ref{eq:thermal}) reduces to
the thermal equation of state of an ideal gas with $2c$ degrees-of-freedom, 
$\beta=c/\ebar$, and (\ref{eq:S}) gives the energy dependency of the entropy of an 
ideal gas, $S=c\log\ebar$. Superstatistics
supposedly enters when Beck and Cohen identify $B$ with the Tsallis distribution where $c=1/(q-1)$
and $q$ is the exponent in the Tsallis entropy. The authors write 
$B=\exp\left\{-c\log(1+b\ebar)\right\}$ and expand it for small $b\ebar$. Yet their 
resulting expansion is \textbf{not} in powers of $b\ebar$! The parameter $b=\beta_0/c=
\beta_0(q-1)$, which in our notation is $b=1/E_0$, the inverse of the total energy, does 
not contain the Tsallis exponent $q$. The  expansion is
\begin{eqnarray*}
(1+bE)^{-c} & = & \sum_{k=c}^\infty\,\left(\begin{array}{c}k-1\\c-1
\end{array}\right)(-bE)^{k-c}\\
& = & 1-\beta_0E+\frac{(\beta_0E)^2}{2!}-
\frac{(\beta_0E)^3}{3!}+\cdots\\
& &+(q-1)\frac{(\beta_0E)^2}{2!}-
(2q+1)(q-1)\frac{(\beta_0E)^3}{3!}+\cdots
\end{eqnarray*}
and not their eqn (14) in which $e^{-\beta_0E}$ can be factored out of the sum. This
has fatal consequences on their conclusion in eqn (25) that the Tsallis entropy exponent
$q=\left<\beta^2\right>/\left<\beta\right>^2$, 
which already looked suspicious because it rules out values of $q<1$.
\par
The log-normal pdf for the inverse temperature had already been derived in ref 
\cite{Lavenda} eqn (4.115) where there is a $1/T_i$ missing. This term is due to the
Jacobian of the transformation on going from (4.114) to (4.115). It was derived from an error law for the temperature for
which the geometric mean is the most probable value of the temperature measured. The
geometric mean value of the temperature is the lowest attainable temperature when two bodies
at different initial temperatures are placed in thermal contact, where the processes of 
heat withdrawal and injection and conversion into work are carried out reversibly. Again
there is neither mention---nor need---of superstatics. \par
We have left Beck and Cohen's first example for last because it illustrates two general
principles. Firstly, Beck and Cohen choose a uniform prior pdf in an interval from $a$ to 
$a+b$. According to Jeffreys' first rule \cite{Jeffreys}, a uniform prior is appropriate
when the parameter involved can conceivably assume all values from $-\infty$ to 
$+\infty$. Since the temperature is non-negative, $\beta$ can assume values from $0$ to $\infty$, and for
such a parameter Jeffreys suggests taking its logarithm uniform, (\ref{eq:Jeffreys}).
\par
Secondly, we have always substituted the Legendre transform for the Laplace integral, and this
is not justified in this case because it is not a thermodynamic system with a large 
number of degrees-of-freedom. Consider the following Laplace type integral
\begin{equation}B(E)=\int_a^{a+b}\,e^{-\beta E}\beta^m\,d\beta.\label{eq:B}
\end{equation}
Under the change of variable $\beta=m\gamma$ the integral becomes
\[B(E)=m^{m+1}\int_{a^{\prime}}^{a^{\prime}+b^{\prime}}\,
e^{-m\phi(\gamma)}\,d\gamma,\]
where $a^{\prime}=a/m$, $b^{\prime}=b/m$, and
\[\phi(\gamma)=\gamma E-\log\gamma.\]
The function $\phi$ has a minimum at $\hat{\gamma}=E^{-1}$, or equivalently, 
$\hat{\beta}=m/E$, which is the thermal equation of state for a system of $2m$ 
degrees-of-freedom. Expanding $\phi(\gamma)$ up to second-order in the small difference
$(\gamma-\hat{\gamma})$ gives
\[B(E)\approx m^{m+1}e^{-m-(m+1)\log E}\sqrt{\frac{2}{m}}
\int_{-\sqrt{(m/2)}(1-a/\hat{\beta})}^{\sqrt{(m/2)}[(a+b)/\hat{\beta}-1]}\,
e^{-\tau^2}\,d\tau,\]
where we have introduced the further change of variable, $\tau=\sqrt{(m/2)}
\gamma^{-1}(\gamma-\hat{\gamma})$.\par We now appreciate that no matter how small $b$ is, $m$ can be taken so
large that the value of the integral is changed only slightly when the limits of 
integration are replaced by $\pm\infty$ \cite{Bleistein}. Thus we finally have
\[B(E)\approx\frac{m^me^{-m}\sqrt{2\pi m}}{E^{m+1}}.\]
For large $m$ the numerator is Stirling's approximation to $m!$ up to order $1/m$;
hence, the final expression for the entropy is
\[S(E)=\log\left(\frac{E^m}{m!}\right).\]
\par
Therefore, we can safely use the Legendre transform in place of evaluating the Laplace
transform for values of $m$ for which Stirling's approximation holds. Only in this case
can the principles of statistical thermodynamics be used \cite{Lavenda}. This does not
apply to Beck and Cohen's uniform distribution. Finally, (\ref{eq:B}) is essentially
the Gamma pdf (\ref{eq:G}) in the case of large $b$ so the normalization condition 
(\ref{eq:norm}) is not relevant.

\end{document}